\documentclass[aip,
 prl,%
 amsmath,amssymb,
 reprint,%
twocolumn]{revtex4-2}
\usepackage{graphicx}
\graphicspath{{./pic}}
\usepackage{dcolumn}
\usepackage{bm}
\usepackage{siunitx}
\usepackage{amsmath}
\usepackage{mhchem}

\begin{document}


\title[Top-loading VSM probe for a VTI]{Design and performance of a top-loading vibrating-sample magnetometer for a helium-flow variable-temperature insert}

\author{Alfons G. Schuck}
\affiliation{Institute of Physics, Goethe University Frankfurt, Max-von-Laue Street 1, Frankfurt am Main 60438, Germany}
\author{Sebastian Kölsch}
\affiliation{Institute of Physics, Goethe University Frankfurt, Max-von-Laue Street 1, Frankfurt am Main 60438, Germany}
\author{Jörg Franke }
\affiliation{Institute of Physics, Goethe University Frankfurt, Max-von-Laue Street 1, Frankfurt am Main 60438, Germany}
\author{Michael Huth}
\email[]{michael.huth@physik.uni-frankfurt.de}
\affiliation{Institute of Physics, Goethe University Frankfurt, Max-von-Laue Street 1, Frankfurt am Main 60438, Germany}

\date{\today}

\begin{abstract}
We present a custom-built top-loading probe that adds vibrating-sample magnetometry to an existing helium-flow variable-temperature insert without replacing its cryogenic or magnetic-field infrastructure. The modular instrument combines a detachable sealed drive head, a voice-coil resonator operating near \qty{44}{Hz}, continuous monitoring of the peak vibration amplitude, motorized axial centering, and a compact first-order pickup-coil gradiometer. A numerical flux-linkage calculation reproduces the position and shape of the measured axial response for several vibration amplitudes after multiplication by a common factor of 1.08. Measurements on Ni reference samples demonstrate operation down to \qty{5}{K} and yield effective geometrical factors between \qtyrange{24.0}{31.4}{\volt\second\per\ampere\per\cubic\meter}. At fixed field, the observed voltage scatter corresponds to moment-equivalent one-standard-deviation repeatabilities of \qtyrange{16.5}{27.8}{nA.m^2}. For a \qty{31.7}{nm} Ni film measured at \qty{250}{K}, magnetization curves obtained at three vibration amplitudes approximately collapse after amplitude correction and yield a saturation magnetization of approximately \qty{409}{kA/m} after subtraction of the linear high-field background. The results demonstrate a practical retrofit approach for adding bulk and thin-film magnetometry to VTIs with comparable bore and flange constraints.
\end{abstract}

\maketitle

\section{Introduction}
Vibrating-sample magnetometry was introduced by Foner in 1959 \cite{foner_versatile_1959}. In this method, a magnetized sample is vibrated periodically with respect to a set of pickup coils, thereby inducing a voltage that is proportional to its magnetic moment. The original instrument already detected changes in magnetic moment of the order of \(10^{-5}\) to \(10^{-6}\,\mathrm{emu}\). Subsequent developments improved the pickup-coil geometry and reduced the sensitivity to small positioning errors. In particular, a broad saddle point in the axial response was shown to be advantageous \cite{Bowden_detection_1972}, and the effects of coil geometry, sample dimensions, and the sample holder on the detected signal were investigated systematically \cite{guy_simple_1976,pacyna_general_1982,pacyna_general_1984,zieba_detection_1982}.

The relative simplicity of the measurement principle has motivated numerous custom-built instruments tailored to specific experimental requirements. Examples include compact printed-circuit-board pickup coils for restricted pole gaps \cite{lopez-dominguez_simple_2018} and signal detection based on audio components as an alternative to a dedicated lock-in amplifier \cite{sankhi_low-cost_2020}. Extending a VSM to cryogenic operation introduces additional constraints, including limited space in the cryostat, reliable sample thermalization, gas-flow-induced motion, temperature-dependent mechanical properties, and reproducible positioning of the vibrating sample.

Custom-built VSMs for cryogenic temperatures and high magnetic fields have been reported for several decades. Mangum and Thornton described a compact instrument for very low temperatures and high applied fields \cite{mangum_vibrating_1970}, and Oliveira and Foner demonstrated the operation of a conventional VSM at \(^{3}\)He temperatures \cite{oliveira_operation_1972}. Gerber \textit{et al.} later presented a simple and relatively inexpensive homemade VSM that supported long experiments at liquid-helium temperatures in a superconducting solenoid \cite{gerber_simple_1982}. A further custom design combined a liquid-nitrogen flow cryostat with a room-temperature electromagnet using locally available components \cite{niazi_precision_2000}. These developments show that cryogenic VSM operation is not restricted to commercial platforms. Nevertheless, retrofitting a VSM to an existing variable-temperature insert remains technically demanding, particularly when the available bore is narrow and sample exchange, thermalization, and stable vibration must be combined in a single probe.

Here, we present a custom-built top-loading VSM probe developed as a retrofit for an existing helium-flow variable-temperature insert. The design combines a detachable sealed head, a voice-coil-driven resonator, continuous monitoring of the vibration amplitude, motorized axial sample positioning, and a compact first-order pickup-coil gradiometer. It is intended for measurements on bulk and thin-film samples without replacing the existing cryogenic infrastructure. The probe described here was developed for a Scientific Magnetics cryostat fitted with an in-house VTI. Its design principles can be adapted to systems with comparable bore and flange geometries.

\section{Measurement principle}
VSMs are induction-based magnetometers. In principle, a single pickup coil is sufficient to detect the signal from a vibrating magnetic sample. In practical instruments, however, two or more pickup-coil sections are commonly connected in series with defined winding directions in order to suppress voltages induced by spatially uniform external fields. Two nominally identical, axially separated coil sections connected in opposition form a first-order axial gradiometer: their responses to a uniform field cancel, whereas the leading response is determined by the axial field gradient. A typical second-order axial gradiometer comprises three coil sections with relative winding weights \(+1,-2,+1\), thereby canceling both the spatially uniform field component and its linear axial variation. The present instrument employs two oppositely connected pickup coils and therefore constitutes a first-order axial gradiometer.

The magnetic flux linkage \(\Psi(z)\) is the signed sum of the magnetic fluxes through all individual turns of the series-connected pickup coils when the center of the sample is at the axial position \(z\). Here, \(k\) labels the pickup-coil sections, \(N_k\) is the number of turns in section \(k\), and \(A_{kj}\) is the oriented surface spanned by its \(j\)-th turn. The flux linkage is

\begin{equation}
    \Psi(z)=\sum_k\sum_{j=1}^{N_k}
    \int_{A_{kj}}\vec{B}_{\mathrm{s}}(\vec{r};z)\cdot d\vec{A},
    \label{eqn:flux-linkage}
\end{equation}

where \(\vec{B}_{\mathrm{s}}\) is the magnetic flux density generated by the sample. The signs of the oriented surfaces account for the opposing winding directions of the two pickup coils. According to Faraday's law, their open-circuit induced voltage is

\begin{equation}
    U_{\mathrm{ind}}(t)=-\frac{d\Psi}{dt}.
    \label{eqn:faraday}
\end{equation}

The dependence of the flux linkage on the magnetization distribution can be expressed compactly using the electromagnetic reciprocity principle \cite{hoult_signal-to-noise_1976,zieba_detection_1982}. Let \(\vec{B}_{\mathrm{c}}(\vec{r})\) be the magnetic field that would be generated at \(\vec{r}\) by a hypothetical current \(I_{\mathrm{c}}\) flowing through the complete pickup-coil circuit, including its actual winding directions, and define the coil sensitivity

\begin{equation}
    \vec{s}(\vec{r})=\frac{\vec{B}_{\mathrm{c}}(\vec{r})}{I_{\mathrm{c}}}.
    \label{eqn:coil-sensitivity}
\end{equation}

If \(\vec{r}'\) is a coordinate fixed to the sample and \(V_{\mathrm{s}}\) is its volume, reciprocity gives

\begin{equation}
    \Psi(z)=\int_{V_{\mathrm{s}}}
    \vec{M}(\vec{r}')\cdot
    \vec{s}(\vec{r}'+z\vec{e}_z)\,dV'.
    \label{eqn:reciprocity-flux}
\end{equation}

Thus, every volume element contributes according to its local magnetization and its position within the spatial sensitivity profile of the pickup coils. Assuming that the magnetization remains constant during one vibration period, differentiation of Eq.~\ref{eqn:reciprocity-flux} yields

\begin{equation}
    U_{\mathrm{ind}}(t)=-\dot{z}(t)
    \int_{V_{\mathrm{s}}}\vec{M}(\vec{r}')\cdot
    \frac{\partial\vec{s}}{\partial z}
    (\vec{r}'+z(t)\vec{e}_z)\,dV'.
    \label{eqn:induzierte-spannung}
\end{equation}

For a homogeneously magnetized sample with a fixed magnetization direction \(\vec{e}_M\), we write \(\vec{M}=M\vec{e}_M\) and \(|\vec{m}|=M V_{\mathrm{s}}\). The position-dependent geometrical factor can then be defined explicitly as

\begin{equation}
    g(z)=\frac{1}{V_{\mathrm{s}}}
    \int_{V_{\mathrm{s}}}\vec{e}_M\cdot
    \frac{\partial\vec{s}}{\partial z}
    (\vec{r}'+z\vec{e}_z)\,dV'.
    \label{eqn:geometrical-factor}
\end{equation}

This definition shows that \(g\) contains not only the geometry and winding configuration of the pickup coils but also the position, shape, size, and orientation of the sample. Consequently, samples with different geometries generally require different calibration factors.

The sample position is modulated sinusoidally according to

\begin{equation}
    z(t)=z_0+\hat{z}\sin(\omega_s t+\phi),
    \qquad
    \dot{z}(t)=\omega_s\hat{z}\cos(\omega_s t+\phi),
    \label{eqn:sample-motion}
\end{equation}

where \(\omega_s=2\pi f_s\). If \(g(z)\) varies negligibly over the vibration path, the peak amplitude of the induced voltage at the fundamental frequency is

\begin{equation}
    \hat{U}_{\mathrm{ind}}=\omega_s\hat{z}\,|\vec{m}|\,|g(z_0)|.
    \label{eqn:conversion-function}
\end{equation}

For a finite vibration amplitude, the lock-in amplifier measures the fundamental component and an effective geometrical factor must be used,

\begin{equation}
    g_{\mathrm{eff}}(z_0,\hat{z})=
    \frac{1}{\pi}\int_0^{2\pi}
    g(z_0+\hat{z}\sin\theta)\cos^2\theta\,d\theta.
    \label{eqn:effective-geometrical-factor}
\end{equation}

Equation~\ref{eqn:conversion-function} then remains valid after replacing \(g(z_0)\) by \(g_{\mathrm{eff}}\). Throughout this work, \(\hat z\) denotes the peak displacement from the center position. The SR830 reports the RMS amplitude of the fundamental voltage. The experimentally recorded voltage is therefore

\begin{equation}
    U_{\mathrm{LIA}}=\frac{\hat U_{\mathrm{ind}}}{\sqrt{2}}
    =\frac{\omega_s\hat z\,|\vec m|\,|g_{\mathrm{eff}}|}{\sqrt{2}}
    =\sqrt{2}\pi f_s\hat z\,|\vec m|\,|g_{\mathrm{eff}}|.
    \label{eqn:rms-conversion}
\end{equation}

\section{Design}
\begin{figure*}[htpb]
	\centering
	\includegraphics[width = 16 cm]{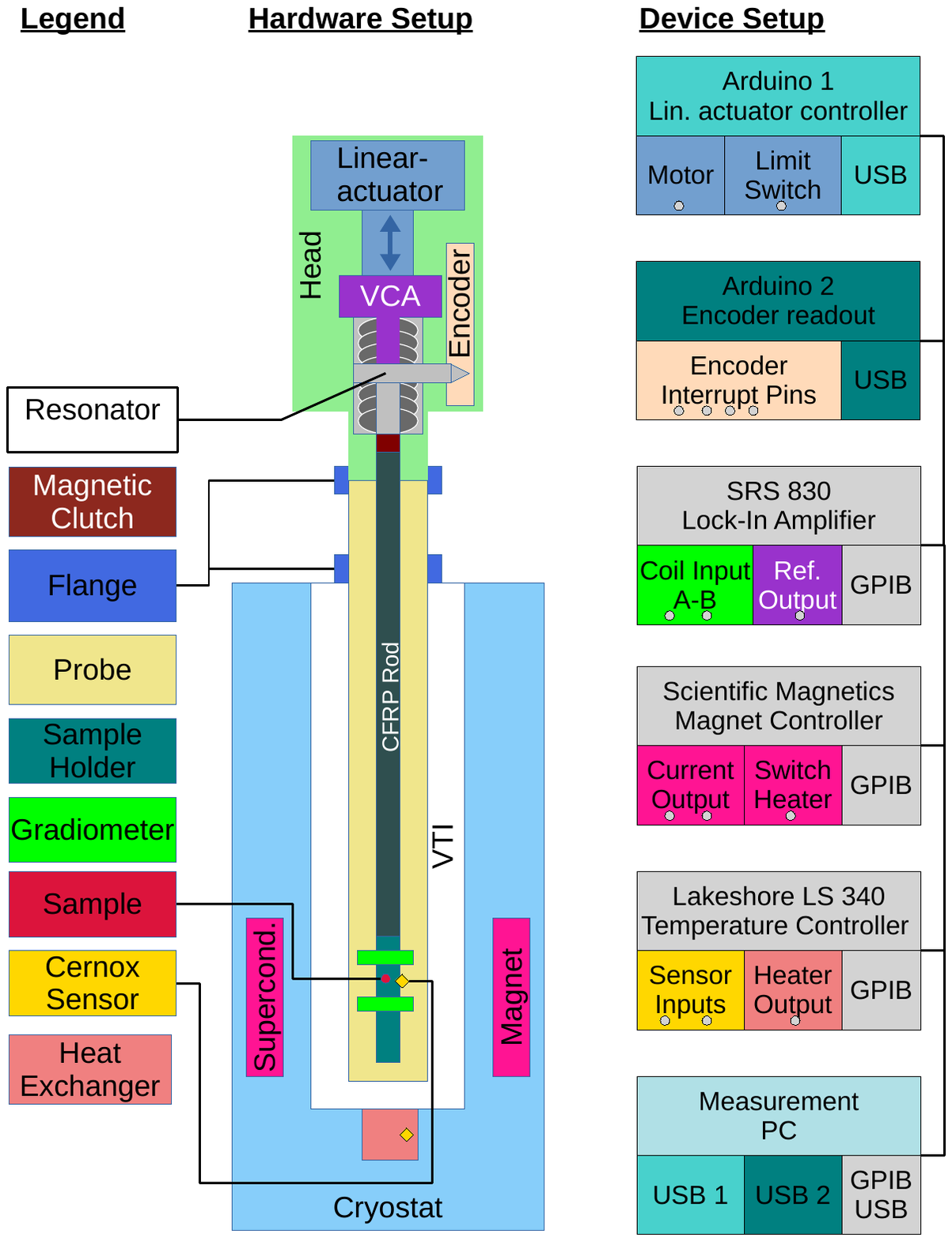}
	\caption{Schematic of the VSM insert and its control electronics. The detachable head contains the positioning and vibration drive, whereas the probe contains the pickup-coil gradiometer, temperature sensor, and sample holder.}
	\label{fig:scheme}
\end{figure*}

The VSM was designed as a top-loading insert for an existing variable-temperature cryostat. It consists of two detachable modules: a head that generates and monitors the axial vibration and a probe that carries the pickup coils, temperature sensor, and sample holder. Figure~\ref{fig:scheme} summarizes the mechanical arrangement and the connections to the control and measurement electronics.

\subsection{Mechanical}

For assembly, the sample is first mounted in the sample holder and attached to the CFRP rod. The probe is then connected to the outer tube, electrically connected, and inserted into the VTI. Finally, the head is mounted on the upper flange and coupled to the vibrating rod. This sequence allows the probe and sample to be handled independently of the heavier head assembly.

\subsubsection{VSM Head} 
\begin{figure}[htpb]
	\centering
	\includegraphics[width = 6cm]{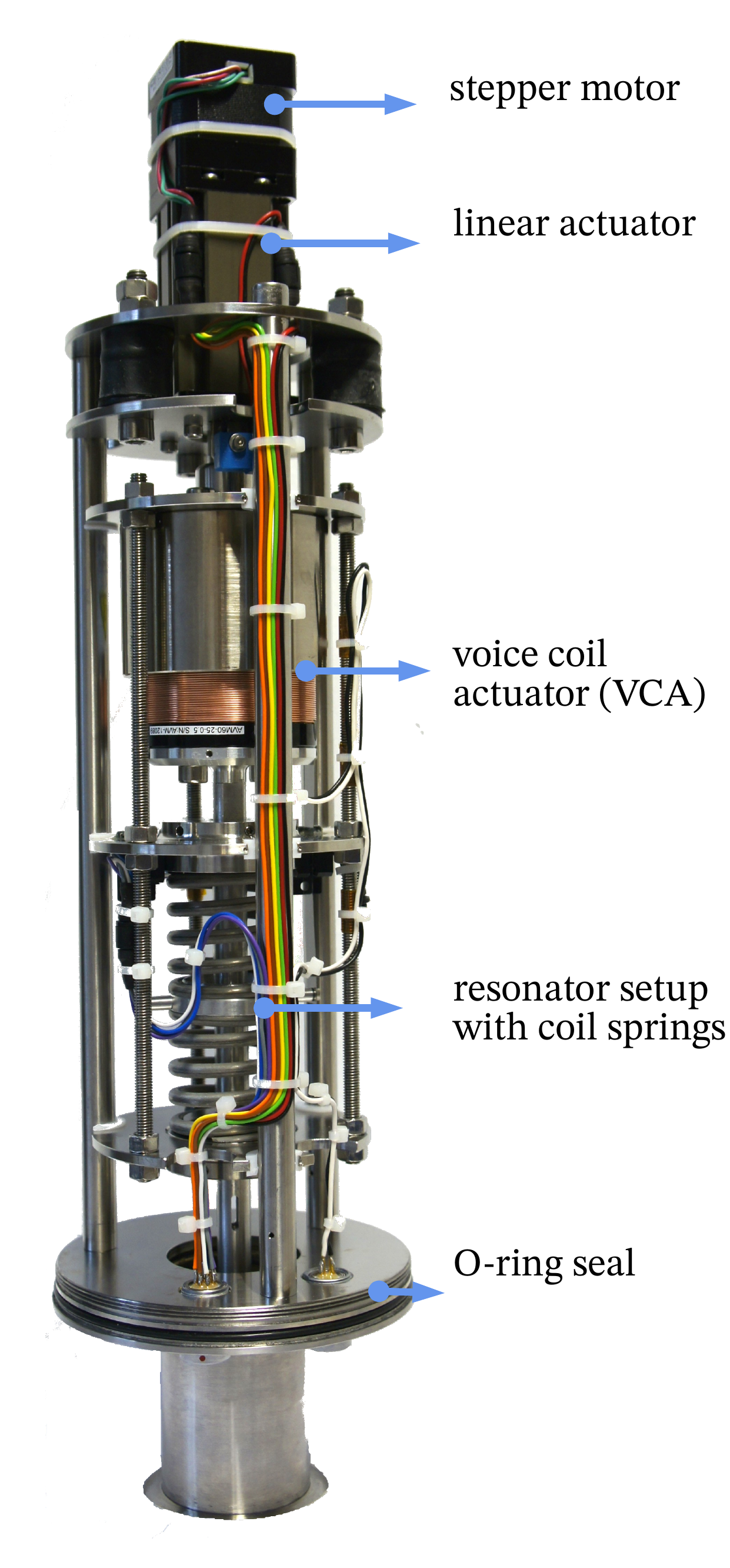}
	\caption{Photograph of the VSM head. The stepper-motor-driven linear actuator sets the axial center position, and the voice coil actuator drives the spring-mounted vibrating assembly. The linear encoder used to measure the vibration amplitude is not visible.}
	\label{fig:vsm_head}
\end{figure} 

The head contains a stepper-motor-driven linear actuator, a limit switch, a voice coil actuator (Akribis AVM60-25), a spring resonator, and a linear quadrature encoder. The linear actuator translates the complete vibrating assembly along the probe axis and is used to center the sample between the pickup coils. The limit switch defines the reference position of this motion.

The voice coil actuator drives a mass-spring system formed by the vibrating assembly and two replaceable compression springs. Its resonance frequency is approximately

\begin{equation}
    f_0=\frac{1}{2\pi}\sqrt{\frac{k_{\mathrm{eff}}}{m_{\mathrm{eff}}}},
    \label{eqn:mechanical-resonance}
\end{equation}

where \(k_{\mathrm{eff}}\) and \(m_{\mathrm{eff}}\) denote the effective spring constant and moving mass, respectively. Springs with a nominal spring constant of approximately \qty{9.9}{N/mm}, a diameter of \qty{30}{mm}, a wire diameter of \qty{3.2}{mm}, and a free length of \qty{43}{mm} were used in the present configuration. The resulting resonance frequency is approximately \qty{44}{Hz}, sufficiently separated from the \qty{50}{Hz} mains frequency. Although the induced voltage increases with frequency at fixed vibration amplitude, the operating frequency must also provide stable mechanical motion at an acceptable drive power.

A CFRP rod provides a lightweight and rigid connection between the resonator and the sample holder. Delrin guide pins suppress lateral motion of the rod. The peak vibration amplitude can reach approximately \qty{5}{mm} and is monitored with a \qty{300}{lpi} linear quadrature encoder (Avago HEDS-9731) connected to an Arduino Uno. Evaluation of both encoder channels and all signal edges corresponds to a nominal displacement increment of approximately \qty{21}{\micro\meter} per count.

The moving rod is coupled magnetically to the probe, while the stationary parts are joined at the flange. The head is enclosed in a removable stainless-steel housing sealed to the VTI, which provides access for maintenance and requires only electrical feedthroughs during operation.

\subsubsection{VSM Probe}
\begin{figure}[htpb]
	\centering
	\includegraphics[width = 6cm]{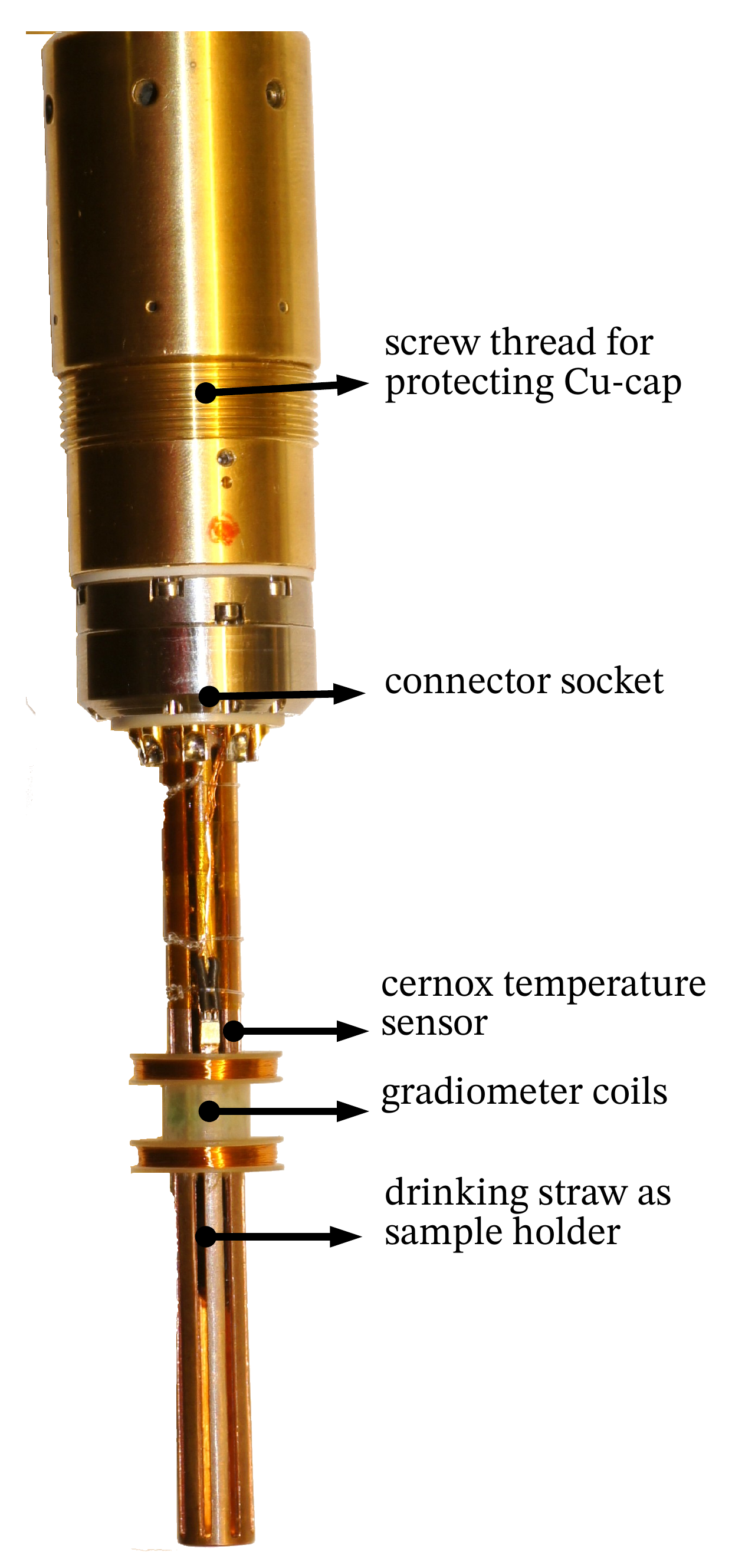}
	\caption{Photograph of the VSM probe without the protective copper cap. The two pickup coils, Cernox temperature sensor, connector, and sample holder are indicated. During operation, a copper cap provides thermal contact to the helium gas and shields the vibrating sample from direct gas flow.}
	\label{fig:vsm_probe}
\end{figure}

The sample holder is a transparent drinking straw with a diameter of \qty{4}{mm}. It is connected to the CFRP rod by a conical, 3D-printed fitting made from transparent Forward AM Ultracur3D RG 35 Rigid Resin (BASF). The fitting is press-fitted into the straw, bonded with GE varnish, and attached to the rod through an M3 thread. This assembly has been used over the temperature range from room temperature to \qty{2}{K}. Transparent materials were chosen to reduce the risk of magnetic background signals from pigments.

The pickup system consists of two nominally identical coils connected in opposition and therefore forms a first-order axial gradiometer. The inner coil diameter is \qty{9}{mm}, and the outer diameter is approximately \qty{16}{mm}, as constrained by the DN40KF flange and the protective cap. Each coil contains 1000 turns with an axial winding length of \qty{2}{mm}; the center-to-center separation of the coils is \qty{9}{mm}. The coils were wound from 42-AWG Electrisola ML240 enameled copper wire. The bare wire diameter is \qty{0.063}{mm}, and the diameter including the polyimide insulation is \qty{0.075}{mm}. Equal winding numbers and opposing connection polarities suppress voltages induced by spatially uniform fields.

A calibrated Cernox sensor is mounted on a slit copper tube surrounding the sample holder, approximately \qty{10}{mm} from the centered sample. All temperatures reported below are the readings of this sensor. During operation, a  copper cap is thermally connected to the tube. It promotes thermal contact with the helium gas and reduces direct excitation of the sample holder by the gas flow. 

An eight-pin connector provides the electrical interface to the probe. Four pins are assigned to the four-wire temperature measurement and two to the series-connected pickup coils. In Fig.~\ref{fig:vsm_probe}, a black straw is used only to improve visibility. The sample holder extends below the lower pickup coil so that its lower end lies outside the most sensitive region of the gradiometer.

\subsection{Electronics}
Depending on sample geometry and magnetic moment, the pickup voltage spans approximately \qty{100}{\nano\volt} to \qty{200}{\milli\volt}. It is measured with a dual-phase lock-in amplifier (Stanford Research Systems SR830) whose reference oscillator also provides the drive signal. The pickup coils are connected to the differential input of the lock-in amplifier. The reported lock-in voltages are RMS values. Unless stated otherwise, the quantitative measurements used a time constant of \qty{300}{ms} and a filter slope of \(24\,\mathrm{dB/octave}\). The specified input-noise density of the SR830 is \qty{6}{\nano\volt\per\sqrt\hertz}. The effective voltage noise additionally depends on the resulting measurement bandwidth.

The sinusoidal reference output of the lock-in amplifier is amplified by a Velleman P8060-2 audio amplifier and applied to the voice coil actuator. The amplifier has a rated output power of \qty{100}{W}. The continuous drive power used for the present setup does not exceed approximately \qty{13}{W}.

The vibration amplitude can vary with friction, temperature, and \ce{^4He} gas flow and is therefore monitored continuously. The two quadrature channels of the HEDS-9731 encoder are evaluated by an Arduino Uno. With four-edge decoding, the \qty{300}{lpi} encoder provides a nominal displacement increment of approximately \qty{21}{\micro\meter}. Because the serial interface cannot transfer every encoder transition to the measurement computer, the Arduino determines the excursion locally. It counts the quadrature increments over each peak-to-peak excursion and forms a moving average over 60 such values. The stored peak-to-peak count \(N_{\mathrm{pp}}\) is converted during analysis to the peak displacement according to

\begin{equation}
    \hat z=N_{\mathrm{pp}}\left(\frac{1}{2}\right)
    \left(\frac{1}{4}\right)
    \frac{\qty{25.4}{mm}}{300}
    =N_{\mathrm{pp}}\times\qty{0.01058}{mm}.
    \label{eqn:encoder-conversion}
\end{equation}

The factors \(1/4\) and \(1/2\) account for four-edge quadrature decoding and conversion from peak-to-peak excursion to peak amplitude, respectively.

A separate Arduino-based controller operates the stepper-motor-driven linear actuator. Before a magnetic measurement, the sample is moved axially through the pickup-coil response and positioned at its maximum, as described below. This motorized procedure provides more reproducible centering than manual placement of the sample in the straw.

\section{Experimental characterization}
\subsection{Mechanical response, sample centering, and calculated pickup response}
The measurements in this subsection establish the mechanical operating point, the axial sample position, and the spatial response of the pickup system before quantitative calibration. Frequency sweeps were first used to select a stable drive frequency and vibration amplitude. A Ni sphere with a mass of \qty{0.383}{g} and a diameter of \qty{4.3}{mm} was then used as a compact, high-moment test sample to map the axial gradiometer response and determine the center position. Its simple geometry also permits a direct comparison between the measured response and the numerical flux-linkage calculation.

\begin{figure}[htpb]
	\centering
	\includegraphics[width = 8.2cm]{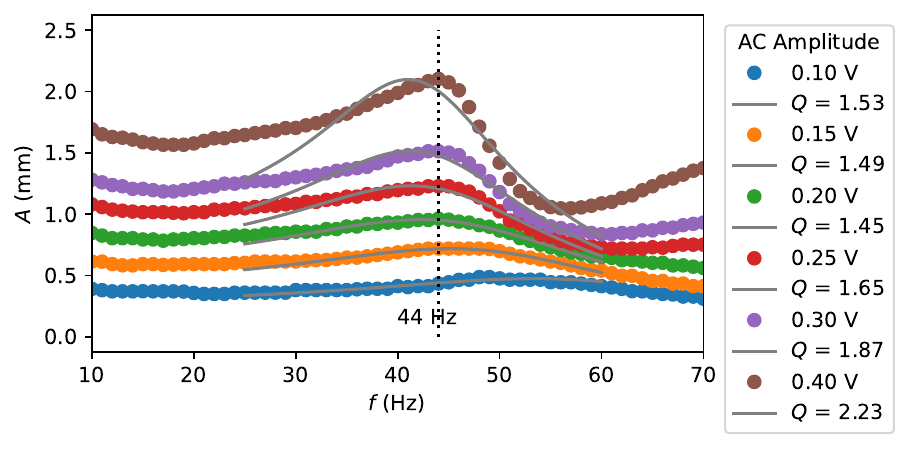}
	\caption{Peak vibration amplitude as a function of drive frequency for the indicated drive voltages. The lines are local fits of Eq.~\ref{eqn:resonator-response} around the principal resonance near \qty{44}{Hz}; the rise at the upper end of the frequency range is not included in the single-mode fit.}
	\label{fig:amplitude}
\end{figure}

The operating frequency was selected from frequency sweeps of the vibration amplitude at fixed drive voltage. As shown in Fig.~\ref{fig:amplitude}, the principal mechanical resonance occurs near \qty{44}{Hz}. The response in the vicinity of this maximum was fitted with a single-mode damped-oscillator model,

\begin{equation}
    A(f)=\frac{a_0}{\sqrt{\left(f_0^2-f^2\right)^2+
    \left(ff_0/Q\right)^2}},
    \label{eqn:resonator-response}
\end{equation}

where \(a_0\) is an amplitude scale, \(f_0\) the resonance frequency, and \(Q\) the mechanical quality factor. For drive voltages between \qty{0.10}{V} and \qty{0.40}{V}, the local fits yield \(Q=1.45\)--\(2.23\); the three lowest drive voltages give the narrower range \(Q=1.45\)--\(1.53\). The broad resonance facilitates stable operation when the mechanical response changes slightly with temperature or helium gas flow, at the expense of a comparatively high drive power. Equation~\ref{eqn:resonator-response} describes only the principal mode. The raw sweeps show a drive-dependent increase above approximately \qty{60}{Hz}, most prominently at \qty{0.40}{V}. This may indicate coupling to an additional mechanical mode or an amplitude-dependent response. Because the measured range does not contain a second maximum, the parameters of a two-mode model cannot be determined independently. The high-frequency increase was therefore excluded from the determination of \(f_0\) and \(Q\).

\begin{figure}[htpb]
	\centering
	\includegraphics[width = 8.2cm]{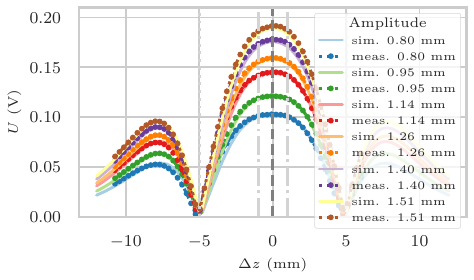}
	\caption{Induced voltage as a function of the axial position of a Ni sphere with a mass of \qty{0.383}{g} and a diameter of \qty{4.3}{mm}, measured at \(\mu_0H=\qty{2}{T}\) for different vibration amplitudes. The lines show the calculated response after multiplication by a common factor of 1.08. The vertical dashed lines indicate the interval within which the measured signal changes by no more than approximately \qty{3}{\percent}. SR830 settings: \qty{300}{ms} time constant and \(24\,\mathrm{dB/octave}\) filter slope.}
	\label{fig:zscan}
\end{figure}

After selection of the operating frequency, the Ni sphere was translated axially through the pickup-coil response at an applied field of \(\mu_0H=\qty{2}{T}\), while the lock-in voltage was recorded (Fig.~\ref{fig:zscan}). This field is well above the saturation field of the sphere. Around the central maximum, a displacement of \(\pm\qty{1}{mm}\) changes the signal by approximately \qty{3}{\percent}. This broad maximum makes the signal comparatively insensitive to small positioning errors. A vibration amplitude of approximately \qty{1}{mm} was subsequently selected as a practical compromise between signal magnitude, stable mechanical motion, and drive power for the samples investigated here.

For the numerical calculation, the 1000 turns of each pickup coil were represented by 21 equally spaced axial sampling planes over the \qty{2}{mm} winding length. An effective loop diameter of \qty{11.4}{mm} was used. At every plane, the sample field calculated with Magpylib \cite{ortner_magpylib_2020} was integrated over the enclosed area using 8001 equal-area points arranged in a Fibonacci pattern \cite{vogel_better_1979}. The opposing winding directions and the \qty{9}{mm} center-to-center coil separation were included when summing the flux linkages. The sample position was varied sinusoidally over one vibration period using 51 time steps, and the induced voltage was obtained from \(-d\Psi/dt\). Demodulation with sine and cosine reference signals normalized by \(\sqrt{2}\) yields the RMS magnitude of the fundamental component, consistent with the SR830 output.

The calculation reproduces the position and overall shape of the measured response for all investigated vibration amplitudes. A common multiplicative factor of 1.08 was applied to the calculated curves in Fig.~\ref{fig:zscan}. Consequently, this comparison tests the spatial response of the gradiometer but does not constitute an independent absolute calibration of its sensitivity.

\subsection{Sample-holder background and short-term voltage repeatability}
\begin{figure}[htpb]
	\centering
	\includegraphics[width = 8.2cm]{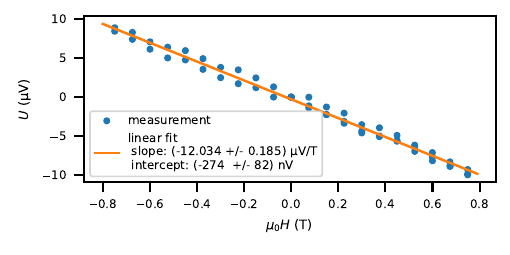}
	\caption{Field-dependent voltage of the drinking-straw sample holder measured with a vibration amplitude of \qty{0.83}{mm}. The line is a linear fit. SR830 settings: \qty{300}{ms} time constant, low-noise reserve mode, and both line-frequency filters enabled. The vibration was allowed to stabilize for \qty{5}{s} before each data point.}
	\label{fig:noise}
\end{figure}

\suppressfloats[t]
For samples with a small magnetic moment, the field-dependent contribution of the drinking-straw sample holder must be determined separately. Figure~\ref{fig:noise} shows the holder signal measured with a vibration amplitude of \qty{0.83}{mm}. Its approximately linear field dependence is consistent with a predominantly diamagnetic background. This reproducible contribution must be subtracted from the sample measurement and is distinct from random voltage fluctuations.

The short-term voltage repeatability was evaluated from \(N=50\) consecutive measurements at fixed field and fixed vibration amplitude. The applied field was held constant throughout each 50-point series. Before each lock-in value was stored, the vibration was switched on and allowed to stabilize for at least ten lock-in time constants plus the mechanical settling time. For the \qty{300}{ms} time constant used here, the total waiting time was \qty{5}{s}. The vibration was switched off after each value had been recorded. No detrending or drift correction was applied; any slow variation during the series is therefore included in the reported scatter. Table~\ref{tab:noise-table} lists the mean voltage \(\bar U\) and its sample standard deviation \(\sigma_U\). The resulting \(\sigma_U\) characterizes the scatter under these specific measurement conditions. It is not by itself a limit of detection for a complete field sweep.

\begin{table}[htpb]
    \centering
    \caption{Short-term voltage repeatability at fixed applied field. Fifty consecutive measurements were evaluated at each field value.}
    \label{tab:noise-table}
    \begin{tabular}{|c|r|r|}
        \hline
        $\mu_0H$ (T)  & $\bar{U}$ ($\mu$V) & $\sigma_U$ (nV) \\ \hline
        0.0   & 0.395   & 40.1   \\ 
        0.1   & 1.314   & 28.2   \\ 
        0.5   & 5.362   & 112.1   \\ \hline
    \end{tabular}

\end{table}
\subsection{Calibration measurements}
Quantitative conversion of the measured voltage into magnetic moment requires a reference with known moment and a comparable sample geometry. Nickel was used for the three reference measurements below. At \qty{298}{K}, its specific saturation magnetization is \(\sigma_{298}=\qty{54.78(15)}{A.m^2/kg}\) \cite{coey_magnetism_2010}. The temperature dependence was calculated using \cite{argyle_deviations_1963}

\begin{equation}
    r(T)=\frac{M(T)}{M(0)}=1-CT^{3/2}-DT^{5/2},
    \label{eqn:ni-temperature}
\end{equation}

with \(C=7.5\times10^{-6}\,\mathrm{K}^{-3/2}\) and \(D=1.5\times10^{-8}\,\mathrm{K}^{-5/2}\). For a sample of mass \(m_{\mathrm{s}}\), the reference moment was calculated as

\begin{equation}
    m(T)=m_{\mathrm{s}}\,\sigma_{298}\frac{r(T)}{r(298\,\mathrm{K})}.
    \label{eqn:ni-reference-moment}
\end{equation}

The effective geometrical factor for each reference geometry and vibration amplitude was calculated from Eq.~\ref{eqn:rms-conversion} as

\begin{equation}
    g_{\mathrm{eff}}=
    \frac{U_{\mathrm{LIA}}}{\sqrt{2}\pi f_s\hat z\,m}.
    \label{eqn:experimental-geometrical-factor}
\end{equation}

\subsubsection{Spherical sample}
The first reference is a Ni sphere, for which the demagnetization factor is \(1/3\). From its mass of \qty{0.383}{g}, Eq.~\ref{eqn:ni-reference-moment} gives a magnetic moment of approximately \qty{22.37}{mA.m^2} at \qty{5}{K}. At \qty{44}{Hz}, a peak vibration amplitude of \qty{1.11}{mm}, and fields above the saturation field of approximately \qty{0.2}{T}, the RMS voltage is approximately \qty{0.15}{V}. The corresponding effective geometrical factor is \(g_{\mathrm{eff,sphere}}=\qty{31.4}{\volt\second\per\ampere\per\cubic\meter}\).

Using \(\sigma_U=\qty{112}{nV}\) measured at \qty{0.5}{T}, the one-standard-deviation moment-equivalent repeatability is

\begin{equation}
    \sigma_m=\frac{\sigma_U}{\sqrt{2}\pi f_s\hat z\,g_{\mathrm{eff}}}
    =\qty{16.5}{nA.m^2}.
    \label{eqn:moment-repeatability}
\end{equation}

This value quantifies short-term scatter under the stated conditions and is not assigned as a general limit of detection.

\begin{figure}[htpb]
	\centering
	\includegraphics[width = 8.2cm]{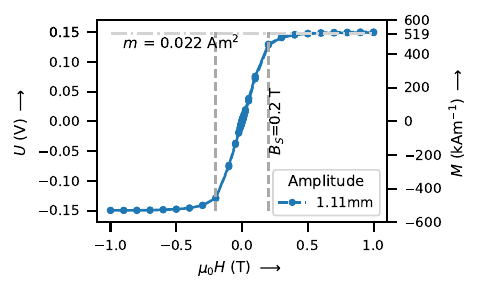}
	\caption{Hysteresis measurement of the Ni sphere with a mass of \qty{0.383}{g} and a diameter of \qty{4.3}{mm}. Measurement conditions: \(f_s=\qty{44}{Hz}\), \(\hat z=\qty{1.11}{mm}\), and \(T=\qty{5}{K}\). The left axis shows the RMS voltage and the right axis the magnetization calculated from the reference moment.}
	\label{fig:ref_sphere}
\end{figure}
\subsubsection{Ni sheet}
The second reference is a Ni sheet with dimensions \(\qtyproduct{4.0 x 4.2 x 0.6}{mm}\), measured with the applied field perpendicular to its plane. The finite sheet has a demagnetization factor close to, but smaller than, the ideal thin-film limit of one. Its mass of \qty{0.093}{g} corresponds to a reference moment of approximately \qty{5.27}{mA.m^2} at \qty{200}{K}. At \(f_s=\qty{44}{Hz}\), \(\hat z=\qty{0.85}{mm}\), and above the saturation field of approximately \qty{0.61}{T}, the RMS voltage is approximately \qty{0.025}{V}. This yields \(g_{\mathrm{eff,OOP}}=\qty{29.3}{\volt\second\per\ampere\per\cubic\meter}\) and a one-standard-deviation moment-equivalent repeatability of \qty{23.0}{nA.m^2} when \(\sigma_U=\qty{112}{nV}\) is used.

\begin{figure}[htpb]
	\centering
	\includegraphics[width = 8.2cm]{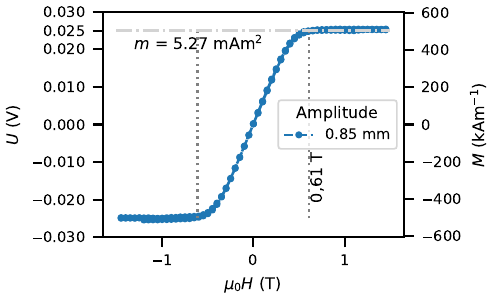}
	\caption{Hysteresis measurement of the Ni-sheet reference in the out-of-plane configuration (\(\qtyproduct{4.0 x 4.2 x 0.6}{mm}\), \qty{0.093}{g}). Measurement conditions: \(f_s=\qty{44}{Hz}\), \(\hat z=\qty{0.85}{mm}\), and \(T=\qty{200}{K}\). The left axis shows the RMS voltage and the right axis the calculated magnetization.}
	\label{fig:ref_sheet_oop}
\end{figure}

The third reference is a Ni sheet with dimensions \(\qtyproduct{5.1 x 5.2 x 0.6}{mm}\), measured with the applied field parallel to its plane. Its mass of \qty{0.151}{g} corresponds to a reference moment of approximately \qty{8.55}{mA.m^2} at \qty{200}{K}. The lower saturation field of approximately \qty{0.1}{T} compared with the out-of-plane configuration is consistent with the smaller in-plane demagnetizing field. At \(f_s=\qty{44}{Hz}\) and \(\hat z=\qty{0.86}{mm}\), the RMS saturation voltage is approximately \qty{34.5}{mV}. The resulting factor is \(g_{\mathrm{eff,IP}}=\qty{24.0}{\volt\second\per\ampere\per\cubic\meter}\), corresponding to a one-standard-deviation moment-equivalent repeatability of \qty{27.8}{nA.m^2} for \(\sigma_U=\qty{112}{nV}\).

\begin{figure}[htpb]
	\centering
	\includegraphics[width = 8.2cm]{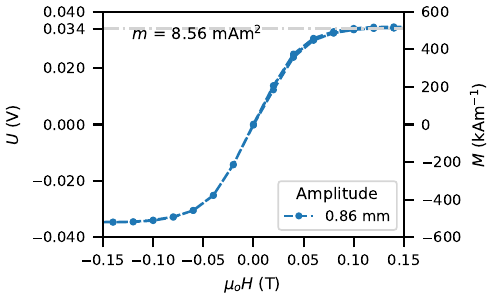}
	\caption{Hysteresis measurement of the Ni-sheet reference in the in-plane configuration (\(\qtyproduct{5.1 x 5.2 x 0.6}{mm}\), \qty{0.151}{g}). Measurement conditions: \(f_s=\qty{44}{Hz}\), \(\hat z=\qty{0.86}{mm}\), and \(T=\qty{200}{K}\). The left axis shows the RMS voltage and the right axis the calculated magnetization.}
	\label{fig:ref_sheet_ip}
\end{figure}

\subsubsection{Thin film sample}
A Ni film was thermally evaporated onto an \(a\)-cut \ce{Al2O3} substrate held at \qty{120}{\celsius}. The base pressure was approximately \qty{5e-6}{hPa}. A quartz-crystal monitor indicated a deposition rate between \qtyrange{0.5}{1}{\angstrom\per\second} over \qty{10}{min}, corresponding to a nominal thickness of approximately \qty{50}{nm}.

\begin{figure}[htpb]
	\centering
	\includegraphics[width = 8.2cm]{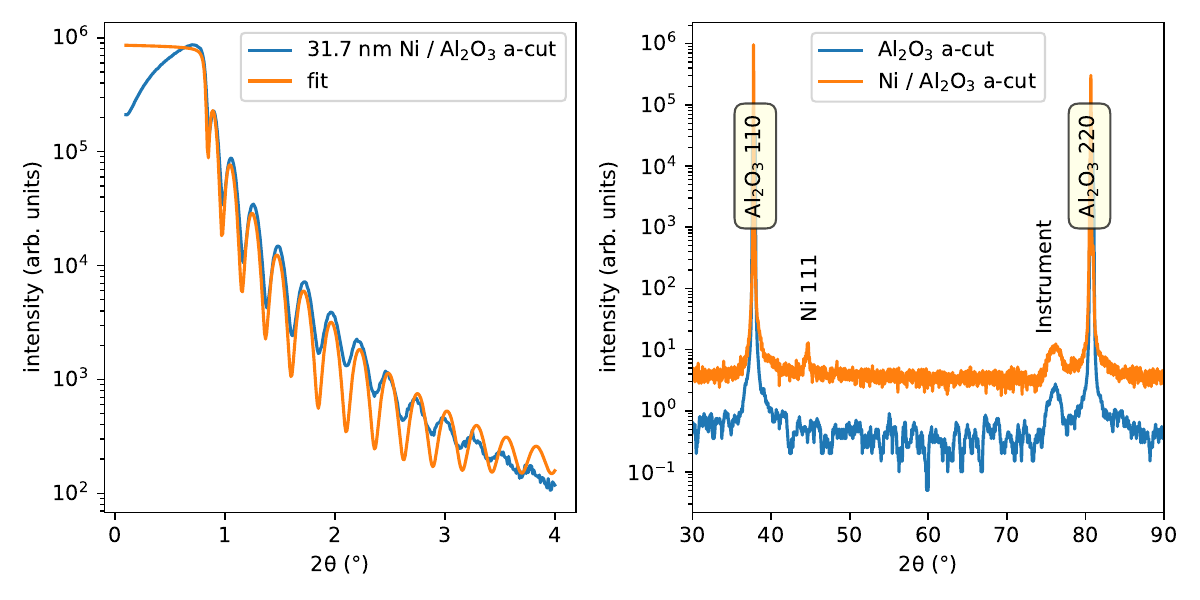}
	\caption{Left: x-ray reflectometry curve of the Ni film and corresponding multilayer fit, yielding a thickness of \qty{31.7(4)}{nm}. Right: symmetric diffraction scans of the Ni film on \ce{Al2O3}(11-20) and a bare \(a\)-cut substrate. The film scan is vertically offset for clarity. The small feature near \qty{78}{\degree} originates from the instrument or sample holder.}
	\label{fig:Ni-XRD}
\end{figure}

The film was characterized with a Bruker D8 Discover diffractometer using monochromatized \ce{Cu_{K,\alpha}} radiation. The x-ray reflectometry curve exhibits Kiessig fringes up to approximately \(2\theta=\qty{4}{\degree}\) (Fig.~\ref{fig:Ni-XRD}). A multilayer fit performed with the Bruker DiffracPlus Leptos software gives a Ni thickness of \(\qty{31.7(4)}{nm}\). This XRR-derived thickness, rather than the nominal quartz-monitor value, was used to calculate the film volume and magnetization. The symmetric diffraction scan confirms the crystalline Ni contribution but is not used for the thickness determination.

\begin{figure}[htpb]
	\centering
	\includegraphics[width = 8.2cm]{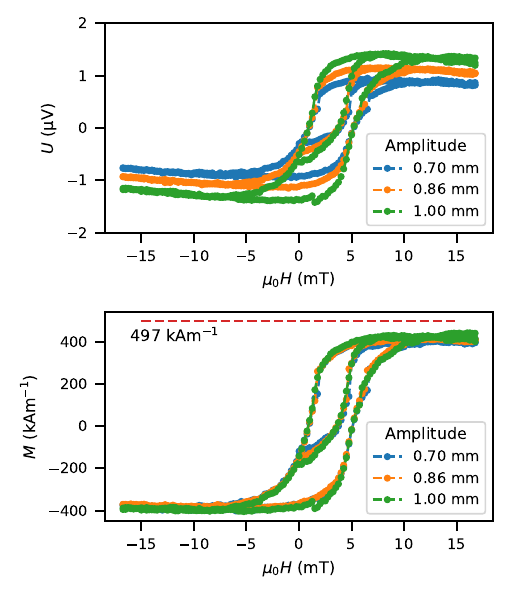}
	\caption{Hysteresis measurements of the \qty{31.7(4)}{nm} Ni film on \(a\)-cut \ce{Al2O3} with lateral dimensions \(\qtyproduct{4.9 x 5.1}{mm}\). Measurement conditions: \(f_s=\qty{44}{Hz}\) and \(T=\qty{250}{K}\). Upper panel: recorded voltage for three vibration amplitudes. Lower panel: magnetization after amplitude correction and subtraction of the linear high-field background attributed to the sample holder and substrate. The dashed line indicates the expected bulk-Ni value at \qty{250}{K}.}
	\label{fig:ni-tf-sample_ip}
\end{figure}

The sample was cut to lateral dimensions of \(\qtyproduct{4.9 x 5.1}{mm}\) and measured at \qty{250}{K} with three vibration amplitudes. Figure~\ref{fig:ni-tf-sample_ip} shows that the raw voltage increases with vibration amplitude, whereas the converted magnetization curves approximately collapse. This provides an internal consistency check of the amplitude correction over the investigated range. For each curve, the approximately linear diamagnetic contribution from the substrate and sample holder was determined by a linear fit in the saturation region and subtracted.

After subtraction of this linear high-field background, the saturation magnetization is approximately \qty{409}{kA/m}, compared with \qty{497}{kA/m} expected for bulk Ni at \qty{250}{K}. The difference of approximately \qty{18}{\percent} may arise from uncertainties in the magnetic volume, background subtraction, calibration, film density, or a magnetically inactive interfacial or surface region. The present measurements do not distinguish between these contributions and therefore do not justify assigning the difference to a specific microscopic mechanism. The result demonstrates the detection of the thin-film hysteresis and the consistency between different vibration amplitudes, but it is not used as an independent validation of the absolute calibration.

\section{Conclusion}
The results establish the top-loading insert as a functional VSM retrofit for a helium-flow VTI with restricted bore and flange geometry. The measured axial response is consistent with the numerical flux-linkage calculation after application of a common factor of 1.08, supporting the modeled gradiometer response and the motorized centering procedure. Calibration with Ni reference samples yields effective geometrical factors between \qtyrange{24.0}{31.4}{\volt\second\per\ampere\per\cubic\meter} and demonstrates operation down to \qty{5}{K}. Under the stated fixed-field conditions, the corresponding moment-equivalent one-standard-deviation repeatabilities range from \qtyrange{16.5}{27.8}{nA.m^2}. They quantify short-term measurement scatter and should not be interpreted as a universal limit of detection.

For the \qty{31.7}{nm} Ni film measured at \qty{250}{K}, the hysteresis curves obtained at three vibration amplitudes approximately collapse after amplitude correction. This agreement provides an internal check of the amplitude normalization and demonstrates the detection of thin-film hysteresis. The derived saturation magnetization of approximately \qty{409}{kA/m}, compared with the bulk-Ni value of \qty{497}{kA/m}, also shows that absolute thin-film magnetization remains sensitive to calibration, background subtraction, magnetic volume, film density, and sample positioning. The same apparatus has also been used for low-temperature VSM measurements of epitaxial \ce{EuPd2} thin films at \qty{5}{K} \cite{schuck_strain-induced_2024}. Taken together, these results demonstrate a modular route for adding bulk and thin-film magnetometry to VTIs with comparable mechanical constraints while making the present metrological limits explicit.

\section*{Acknowledgements}
\paragraph{Funding information}
The authors are grateful for the funding from the Deutsche Forschungsgemeinschaft (DFG, German Research Foundation), Grant No. TRR 288 - 422213477 (Project No. A04).

\bibliographystyle{unsrt} 
\bibliography{vsm_literature.bib}

\end{document}